\documentclass[a4paper,11pt]{article}
\pdfoutput=1 

\usepackage{jcappub} 

\usepackage[T1]{fontenc} 
\usepackage{color}
\definecolor{Red}{rgb}{0.65,0.08,0.05}
\definecolor{Blue}{rgb}{0.05,0.08,0.65}

\newcommand{\dd}{{\rm d}}
\newcommand{\ii}{{\rm i}}

\newcommand{\vx}{\textbf{x}}
\newcommand{\vd}{\textbf{d}}

\newcommand{\vq}{\textbf{q}}
\newcommand{\vk}{\textbf{k}}

\newcommand{\etain}{{\eta_{\rm in}}}

\newcommand{\mH}{{\cal H}}

\newcommand{\mD}{{\cal D}}

\newcommand{\mF}{{\cal F}}
\newcommand{\mS}{{\cal S}}

\newcommand{\mP}{{\cal P}}
\newcommand{\Dirac}{{\delta_{\rm Dirac}}}

\newcommand{\vPi}{{\tau}}
\newcommand{\tP}{{P}}

\newcommand{\tx}{\tilde{x}}

\begin{document}
\title{Cosmological Perturbation Theory for streams of relativistic particles}

\author[a,b]{H\'el\`ene Dupuy,}
\author[b,a]{Francis Bernardeau}

\affiliation[a]{Institut de Physique Th\'eorique, CEA, IPhT, F-91191 Gif-sur-Yvette,\\
CNRS, URA 2306, F-91191 Gif-sur-Yvette, France}
\affiliation[b]
{UPMC - CNRS, UMR7095, Institut d'Astrophysique de Paris, F-75014, Paris, France}

\emailAdd{helene.dupuy@cea.fr}      
\emailAdd{francis.bernardeau@iap.fr}

\abstract{
Motion equations describing streams of relativistic particles and their properties are explored in detail in the framework of Cosmological Perturbation Theory. Those equations, derived in any metric both in the linear and nonlinear regimes, express the matter and momentum conservation. In this context we extend the setup of adiabatic initial conditions - that was initially performed in the conformal Newtonian gauge - to the synchronous gauge. The subhorizon limit of the nonlinear motion equations written in a generic perturbed Friedmann-Lema\^{i}tre metric is then derived and analyzed. We show in particular that the momentum field $P_{i}(x)$ is always potential in the linear regime and remains so at subhorizon scales in the nonlinear regime. Finally the equivalence principle is exploited to highlight invariance properties satisfied by such a system of equations, extending that known for streams of non-relativistic particles, namely the extended Galilean invariance.
}

\date{\today}
\maketitle

\bigskip


\section{Introduction}
Entering in the era of precision cosmology requires an examination of the structure growth in minute detail. So far, the impact of neutrinos on large-scale structure formation has often been overlooked, especially in the nonlinear regime, because of the difficulties encountered when accounting for the gravitational dynamics of such particles is needed (see recent attempts in \cite{2009JCAP...06..017L,2010PhRvD..81l3516S,2014arXiv1408.2995B}). Since neutrinos have been shown to be massive, the situation has evolved: it has become crucial to include neutrino masses in cosmological models to study their impact on the late-time growth of structure. On the one hand, one indeed needs to ensure that the estimation of the fundamental cosmological parameters is not undermined by the presence of massive neutrinos. On the other hand, doing so is an efficient way to improve our knowledge of those enigmatic particles because, as mentioned in many references \cite{2011PhRvD..83d3529S,2012PhRvD..85h1101R,2013JCAP...01..026A,2011JCAP...03..030C,2011arXiv1110.3193L,2009A&A...500..657T}, the signature of neutrino masses on cosmological observables is significant enough for those masses to be constrained observationally.

From a theoretical point of view, these types of investigations are not straightforward. Contrary to Cold Dark Matter (CDM) particles, neutrinos are relativistic at horizon crossing so their velocity dispersion is not negligible and the Newtonian approximation does not hold. The phase-space distribution of neutrinos is therefore difficult to describe. 
Usually, this question is addressed in the linear regime only by performing a harmonic decomposition 
of the phase-space distribution function. As presented in some standard references (e.g. \cite{1994ApJ...429...22M} and its companion paper Ref. \cite{Ma:1995ey}), this approach leads to a hierarchy of equations called the Boltzmann hierarchy. It is possible to examine in full detail the impact of such results on cosmology, as reviewed for instance 
in Ref. \cite{Lesgourgues2006}, but this analysis is limited to the linear regime, which is too restrictive to be in balance with the current surveys. Indeed, observational projects aiming at putting constraints on neutrino masses are all sensitive to the nonlinear growth of structure, at least in the mildly nonlinear regime.  Problematically, the extension into the nonlinear regime of the phase-space harmonic decomposition has proved to be
very cumbersome\footnote{The only attempt we are aware of is described in \cite{vanderijtphd}.}.

An alternative approach, which we further explore here,  has been put forward recently in  Ref. \cite{2014JCAP...01..030D}.
In this study neutrinos, or more generally any non-interacting relativistic or non-relativistic particles, were described as a collection of flows evolving independently from one  another. Such a study takes advantage of the fact that the particles at play are free-streaming. This is this property that allows to replace the standard study of a single multi-stream fluid by that of a collection of single-stream fluids.
Somehow it takes  inspiration of the CDM description (for which the single-stream approximation is an effective approximation, see e.g. \cite{2013arXiv1311.2724B}), the aim being to establish a similar description for neutrinos.  
In our first paper we demonstrated that, at linear level, the method we proposed  led to the same results as the standard one.
In the present article, we explore in more detail the formalism associated with the approach we developed. In particular we derive the motion equations in a more general framework, focusing on the symmetry properties of the resulting equations. Ward identities, resulting from those invariances, are also presented.  What motivates this task is the fact that carrying out Perturbation Theory calculations requires a good understanding of the mode coupling structure.

The organization of the paper is the following. In section \ref{Sect:multiflow} we recall the specificities of a multi-fluid description. Then we derive the fully nonlinear motion equations satisfied by single-stream fluids of relativistic particles. The aim of the following section is to bring out some remarkable properties related to the linear regime. Finally, section \ref{Relativistic PT} explains how to deal with relativistic streams in Perturbation Theory. In particular a global motion equation, fully nonlinear but in which the terms that are subdominant at subhorizon scales have been dropped, is presented in this section. We then present and comment the key properties of this system before concluding and discussing perspectives.

\section{A multi-fluid description of non-interacting relativistic particles}
\label{Sect:multiflow}


We recall here the method we developed to describe a fluid of non-interacting relativistic massive particles as a collection of streams or flows\footnote{Note that, when applied to neutrinos, this construction is valid for any given mass eigenstate. If the masses are not degenerate, it should therefore be repeated for each three eigenstates.} (we will use hereafter the two terms indistinctively).
This approach requires to properly define the phase-space distribution function $f(x^{i},p_{i},\eta)$, where $x^{i}$ are the comoving positions, 
$p_{i}$ the conjugate momenta of $x^{i}$ and $\eta$ the conformal time. 

The key idea is to split a relativistic multi-flow fluid into several flows in order to enter in the field of application of the single-flow approximation. In the absence of shell-crossing, particles having initially the same velocity will continue to do so throughout the cosmological time. Indeed, in that case,  particles that have the same velocity at the same time are particles that are located at the same place so such particles will travel through the same
gravitational potentials. Those sets of particles are thus single-flow fluids. A convenient way to distinguish between the flows 
is to use initial momenta $p_{i}(\eta_{\rm{in}})$ as labels\footnote{In a homogeneous expanding universe, one can easily show that $p_{i}$ is a constant.}, denoted $\vPi_{i}$.  Each single-flow fluid considered in our multi-fluid approach is therefore defined as the collection of all the particles that have at initial time the comoving momentum $\tau_{i}$. 
The time evolution of each flow itself is encoded in a phase-space distribution function $f^{\rm{one-flow}}(\eta,x^{i},p_{i};\vPi_{i})$. One also introduces the momentum field $P_{i}(\eta,x_{i};\vPi)$, which is the value of the momentum of any particle of the flow labelled by $\vPi_{i}$ at time $\eta$ and position $x_{i}$. The phase-space distribution function $f^{\rm{one-flow}}(\eta,x^{i},p_{i};\vPi_{i})$ then reads
\begin{equation}
f^{\rm{one-flow}}(\eta,x^{i},p_{i};\vPi_{i})=
n_{c}(\eta,\vx; \vPi_{i}) \delta_{\rm{D}}(p_{i}-P_{i}(\eta,\vx; \vPi_{i})),
\end{equation}
where $n_{c}$ is the comoving number density of the flow.

Finally, the overall distribution function $f(\eta,x^{i},p_{i})$ is computed by taking all the single-flow fluids previously defined into account:
\begin{equation}
 f(\eta,x^{i},p_{i})=\sum_{\vPi_{i}}{f^{\rm{one-flow}}(\eta,x^{i},p_{i};\vPi_{i})}
=\sum_{\vPi_{i}}{n_{c}(\eta,\vx; \vPi_{i}) \delta_{\rm{D}}(p_{i}-P_{i}(\eta,\vx; \vPi_{i}))}.
 \end{equation}
Assuming that the parameter $\vPi_{i}$ describes a 3D continuous field, the continuous limit of this expression is naturally
\begin{equation}\label{ftot}
 f(\eta,x^{i},p_{i})
 =\int{\dd^{3}\vPi_i}\,n_{c}(\eta,\vx;\vPi_{i}) \delta_{\rm{D}}(p_{i}-P_{i}(\eta,\vx;\vPi_i)).
\end{equation}
In this context, one can see that the integration over phase-space momenta usually performed to compute global physical quantities associated with a single multi-flow fluid is replaced by a sum over the single-flow fluids labelled by $\vPi_i$ (i.e. a sum over all the possible initial momenta or velocities). It implies in particular that, for any functional form $\mF(p_{i})$, we have
\begin{equation}\label{mapping1}
\int{\dd^{3}p_{i}}  \,f(\eta,x^{i},p_{i})\ \mF(p_{i})=  \int{\dd^{3} \vPi_{i}}\,n_c(\eta,\vx;\vPi_{i})\ \mF(P_{i}(\eta,\vx;\vPi_{i})).
\end{equation}

Eq. (\ref{ftot}) shows that, to determine the time evolution of the phase-space distribution function, one needs to study the time evolution of the comoving number densities
and of the momentum fields. A derivation of the corresponding motion equations in the conformal Newtonian gauge was presented in detail in \cite{2014JCAP...01..030D}. In the next section, we succinctly generalize this derivation to an arbitrary spacetime.

\section{Derivation of the nonlinear motion equations of relativistic massive particles}

\subsection{Evolution equation of the comoving number densities}

The comoving number density of a fluid, single-flow or not,  is related to its
phase-space distribution function in a very simple way, 
\begin{equation}
n_{c}(\eta,x^{i})=\int \dd^{3}p_{i}\,f(\eta,x^{i},p_{i}).
\end{equation}
Note that  $n_{c}$ is not necessarily the number density measured by an observer at rest in the metric (see subsection \ref{sec:energymomentum}).
Its evolution equation can be derived straightforwardly from the conservation equation\footnote{The Vlasov equation derives from this equation under the assumption that the motion equations are Hamiltonian.} satisfied by the phase-space distribution function $f$,
\begin{equation}
\frac{\partial}{\partial\eta}f+\frac{\partial}{\partial x^{i}}\left(\frac{\dd x^{i}}{\dd \eta}f\right)
+\frac{\partial}{\partial p_{i}}\left(\frac{\dd p_{i}}{\dd \eta}f\right)=0.
\end{equation}
The only assumption made here is that the particles we consider are conserved (they do not decay because no disintegration or scattering process is at play at the time of interest). 
Integrating over momenta leads to
\begin{equation}
\frac{\partial}{\partial\eta}n_{c}+\frac{\partial}{\partial x^{i}}
\left(\int\dd^{3}p_{i} \frac{\dd x^{i}}{\dd \eta}f\right)=0.
\end{equation}
For a single-flow fluid, $\dd x^{i}/\dd \eta$ can be expressed in terms of momenta. More specifically,
\begin{equation}
\frac{\dd x^{i}}{\dd\eta}=\frac{\dd x^{i}}{\dd\tau}\frac{\dd \tau}{\dd\eta}=\frac{p^{i}}{p^{0}},
\end{equation}
where $\tau$ is the particle proper time and $p^{i}$ and $p^{0}$ are related to $p_{i}$ through the metric and the on-shell mass constraint. 
As a result, for a single-flow fluid we have
\begin{equation}
\frac{\partial}{\partial\eta}n_{c}+\frac{\partial}{\partial x^{i}}
\left(\frac{P^{i}}{P^{0}}n_{c}\right)=0,\label{ncevol}
\end{equation}
with $P^{\mu}(\eta,x^{i})=g^{\mu\nu}P_{\nu}(\eta,x^{i})$ and $P^{\mu}P_{\mu}=-m^{2}$. Note that this motion equation does not rely
on any perturbative expansion of the metric.

\subsection{Evolution equation of the momentum fields}

A fluid in which particles are neither created nor annihilated nor subjected to diffusion, as is the case with the fluids considered here, obeys 
general conservation laws such as
\begin{equation}\label{Jconservation}
T^{\mu\nu}_{\phantom{0}\phantom{0};\mu}=0,
\end{equation}
where $T^{\mu\nu}$ is the energy-momentum tensor
and where we adopt the standard notation ; to indicate a covariant derivative. The conservation of the particles can besides be expressed as
\begin{equation}\label{Jconservation}
J^\mu_{\phantom{0};\mu}=0,
\end{equation}
where $J^\mu$ is the particle four-current. Noting that, for a single-flow fluid,  the energy-momentum tensor $T^{\mu\nu}$ is related to $J^{\nu}$ and to the momentum field $P^\mu$ by 
\begin{equation}
T^{\mu\nu}=-P^\mu J^\nu,
\end{equation}
the conservation of the energy momentum tensor combined with the conservation of the four-current imposes $P^\mu_{\phantom{0};\nu}J^\nu=0$. Moreover, the energy-momentum tensor being symmetric, $T^{\mu\nu}=-P^\mu J^\nu$ gives $J^i=\dfrac{P^i}{P^0}J^0$, whence 
\begin{equation}
P^\nu P_{\mu;\nu}=0.
\end{equation}
Expressing the covariant derivative in terms of the spatial derivative finally gives the following motion equation
\begin{equation}\label{GeneralPiEvol1}
P^\nu P_{\mu,\nu}=\dfrac{1}{2} P^\sigma P^\nu g_{\sigma \nu,\mu}.
\end{equation}
Together with eq. (\ref{ncevol}), it  dictates the time evolution of a collection of massive relativistic particles evolving in an arbitrary metric $g_{\alpha \beta}$. Once again, this equation has been obtained without performing any perturbative expansion of the metric. Besides, the time coordinate of the momentum field being related to the spatial ones thanks to the on-shell mass constraint, one can restrict the coordinates $\mu$ to spatial coordinates $i$ in the previous equation,  
\begin{equation}
P^\nu P_{i,\nu}=\dfrac{1}{2}P^\sigma P^\nu  g_{\sigma \nu,i}.\label{GeneralPiEvol}
\end{equation}
Eq. (\ref{GeneralPiEvol}) can therefore be considered as the second motion equation governing the time evolution of the flow in the nonlinear regime.

\subsection{Explicit form for the momentum field in a generic perturbed Friedmann-Lema\^{i}tre metric}

As an illustration, we present in this section the explicit form of the motion equation satisfied by the momentum field in a generic perturbed Friedmann-Lema\^{i}tre metric. The metric we use reads

\begin{equation}
\dd{s^2}=a^2(\eta)\left[-(1+2A)\dd{\eta^2}+2B_i\dd{x^i}\dd{\eta}+(\delta_{ij}+h_{ij})\dd{x^i}\dd{x^j}\right],
\end{equation}
where $\eta$ is the conformal time,  $x^i$  $( i = 1, 2, 3)$ are the Cartesian spatial comoving coordinates, $a(\eta)$ is the scale factor and $A$, $B_i$ and $h_{ij}$ are respectively the time-time, time-space and space-space metric perturbations. 
Units are chosen so that the speed of light in vacuum is equal to unity and the expansion history of the universe, encoded in the time dependence of $a$, is driven by the overall matter and energy content of the universe. The equation of motion for the momentum is then
\begin{equation}\label{Deriv tot P_i 1}
\dfrac{\dd P_i}{\dd \eta}=a^2(\eta)\left[-P^0\partial_i A+P^j\partial_i B_j+\dfrac{1}{2}\dfrac{P^jP^k}{P^0}\partial_ih_{jk}\right],
\end{equation}
where we define the operator $\dd/\dd\eta$ as
\begin{equation}
\frac{\dd}{\dd\eta}\equiv \frac{\partial}{\partial\eta}+\frac{P^{i}}{P^{0}}\frac{\partial}{\partial x^{i}}.
\end{equation}

This equation is, on the one hand, a relativistic generalization of one of the two equations describing CDM and, on the other hand, a generalization to an arbitrary perturbed Friedmann-Lema\^{i}tre metric of eq. (2.13) of \cite{2014JCAP...01..030D},
which corresponds to the conformal Newtonian gauge (in which, by definition, $A=\psi$, $B_i=0$ and $h_{ij}=-2 \phi \delta_{ij}$).
For example, to write the motion equation in the synchronous gauge, one only has to set $A=0$ and $B_i=0$ in eq. (\ref{Deriv tot P_i 1}).
Furthermore, in the non-relativistic limit, we simply have in the synchronous and conformal Newtonian gauges $P^{0}=m/a (1-A)$ and $P^{i}\ll m/a$ so that eq. (\ref{Deriv tot P_i 1}) simply takes the form 
\begin{equation}\label{DerivNonRel}
\dfrac{\dd P_i}{\dd \eta}=-a m\partial_i A.
\end{equation}
It is of course nothing but Newton's second law written using the conformal time.

\subsection{Link with the Einstein equations}
\label{sec:energymomentum}
In general, the energy-momentum tensor is related to the phase-space distribution function through
\begin{equation}
T_{\mu\nu}(\eta,x^{i})=\int\dd^{3}p_{i}\,(-g)^{-1/2}\frac{p_{\mu}p_{\nu}}{p^{0}}\,f(\eta,x^{i},p_{i}).
\end{equation}
When considering single-flow fluids, integration over momenta can be done straightforwardly and the resulting expression is
\begin{equation}
T^{\rm one-flow}_{\mu\nu}=\frac{P_{\mu}P_{\nu}}{(-g)^{1/2}P^{0}}n_{c}.
\end{equation}
This result is important since it gives the contribution of each flow to the Einstein equations.  The particle four-current $J_{\mu}$ can also
be computed very easily, 
\begin{equation}
J^{\rm one-flow}_{\mu}=-\frac{P_{\mu}}{(-g)^{1/2}P^{0}}n_{c}.
\end{equation}
Besides, the expression of the particle four-current can be used to express the number density of particles as seen by
an observer at rest in the metric, denoted $n$. If one calls the four-velocity  of such an observer $U^{\mu}$, one indeed has
\begin{equation}
n=U^{\mu}J_{\mu}.
\end{equation}
Admitting that the four-velocity of such an observer satisfies $U^{i}=0$, one gets
$U^{0}=(-g_{00})^{-1/2}$ from the constraint $U^{\mu}U_{\mu}=-1$, whence 
\begin{equation}
n=-\frac{P_{0}}{(-g_{00})^{1/2}(-g)^{1/2}P^{0}}\,n_{c}.
\end{equation}
In Ref.  \cite{2014JCAP...01..030D}, the evolution equation of the number density of neutrinos was formulated in terms of $n$ rather than $n_{c}$ but
the two approaches are of course equivalent. Those equations allow to study explicitely the Einstein equations after recombination because, when all the cosmic components are free-streaming, one can write
\begin{equation}
G_{\mu\nu}(\eta,x^{i})=8\pi G\sum_{\rm species\ and\ flows}T^{\rm one-flow}_{\mu\nu}(\eta,x^{i}),
\label{EEquations}
\end{equation}
the sum being performed over all species participating in the cosmic fluid and $G$ being Newton's constant. Indeed, the formalism we are proposing  can be applied to relativistic as well as to non-relativistic species so no contribution is missing. Together with the evolution equation of $n_{c}$ and $P_{i}$, these equations form a closed set of equations.

In the following we analyze this system in more detail, starting with the study of its 
invariance properties.

\subsection{Invariance properties}\label{T1}
In this subsection, we present transformation laws that leave the motion equations unchanged. 

A priori, a whole set of transformation
laws can be derived from a change of variables of the form
\begin{equation}
{x}^\mu\ \to\ \tilde{x}^\mu(x^{\mu}).
\end{equation}
In this study, we are interested in generalizing the so-called extended Galilean invariance that the standard motion equations of non-relativistic  particles  (or, equivalently, the motion equations of this paper written in the non-relativistic limit) satisfy. The corresponding transformation laws are explicitly given by the following changes of coordinates and fields (see e.g. \cite{2013NuPhB.873..514K}\footnote{In this study several particular cases, corresponding to different possible time dependencies of the spatial translation, are explored.}),
\begin{eqnarray}
\tilde{x}^{i}&=&x^{i}+d_{i}(\eta),\ \tilde\eta=\eta,\\
\tilde{V}^{i}&=&V^{i}+\frac{\dd}{\dd \eta}d_{i}(\eta),\ \tilde\delta=\delta,\\
\tilde{A}&=&A-\mH \frac{\dd}{\dd \eta}d_{i}(\eta)x^{i}- \frac{\dd^{2}}{\dd \eta^{2}}d_{i}(\eta)x^{i},
\end{eqnarray}
where we use the notations of this paper and where $d_{i}(\eta)$ is an arbitrarily time-dependent uniform field, $V^i$ is the velocity field and $\delta$ is the density contrast. 
Such an invariance corresponds to an extension of the Galilean invariance in the sense that the displacement can depend on time. It actually derives from the equivalence principle.
As pointed out in recent studies, such as \cite{2013arXiv1311.2724B}, the extended Galilean invariance plays a significant role in the computation of correlation functions involved in the study of large-scale structure formation. To generalize it,  
we will explore in the following the consequences of coordinate transforms closely connected to special Lorentz transformations.

So let us consider a transformation $x^{\mu}\to\tilde{x}^{\mu}$ defined so that
\begin{equation}\label{trans dx}
\dd \tx^{\mu}=\xi^{\mu}_{\nu}\ \dd x^{\nu}.
\end{equation}
It is an acceptable transformation provided that
\begin{equation}
\xi^{\mu}_{\nu,\sigma}=\xi^{\mu}_{\sigma,\nu}.\label{diffproperty}
\end{equation}
In particular, it is easy to check that

\begin{eqnarray}
\xi^{0}_{0}&=1+x^{i}\frac{\partial}{\partial\eta}v_{i}(\eta),\ \ \ \ 
\xi^{0}_{i}&=v_{i}(\eta),\label{CoordinateTransform1}\\
\xi^{i}_{0}&=v_{i}(\eta)+u_{i}(\eta),\ \ \ \ 
\xi^{i}_{j}&=\delta^{i}_{j},\label{CoordinateTransform2}
\end{eqnarray}
where $v_{i}(\eta)$ and $u_{i}(\eta)$ are two arbitrarily time-dependent  functions, obeys the property (\ref{diffproperty}).  In the rest of the paper, we will assume that both $v_{i}$ and $u_{i}$
are infinitesimal quantities and we will restrict all calculations to linear order in these quantities.
Besides, the differential operators associated with the new coordinates are related to the previous ones thanks to
\begin{align}
\frac{\partial}{{\partial}{\tilde\eta}}=\left(1-\dot{v}_ix^i\right)\frac{\partial}{{\partial}{\eta}}-\left(v_i+u_{i}\right)\frac{\partial}{{\partial}{x^{i}}} \ \hbox{ and } \
\frac{\partial}{{\partial}{\tilde{x}^{i}}} =-v_i\frac{\partial}{{\partial}{\eta}}+\frac{\partial}{{\partial}{x^{i}}}.
\end{align}
The momentum components therefore transform as (once again $v_{i}$ and $u_{i}$ are kept at linear order)
\begin{eqnarray}
\tilde{P}^{0}&=&(1+x^{i}v_{i,0})P^{0}+v_{i}P^{i}\\
\tilde{P}^{i}&=&(v_{i}+u_{i})P^{0}+P^{i}\\
\tilde{P}_{0}&=&(1-x^{i}v_{i,0})P_{0}-(v_{i}+u_{i})P_{i}\\
\tilde{P}_{i}&=&-v_{i}P_{0}+P_{i},
\end{eqnarray}
the transformation of the comoving numerical density field is given by
\begin{equation}
\tilde{n}_{c}=n_{c}\left(1+v_{i}\frac{P^{i}}{P^{0}}\right)
\end{equation}
and finally the potentials transform the following way,
\begin{eqnarray}
\tilde{A}=A-\mH v_{i}x^{i}-v_{i,0}x^{i},\ \ \ \ 
\tilde{B}_{i}=B_{i}-u_{i},\ \ \ \ 
\tilde{h}_{ij}=h_{ij}-2 \mH\delta_{ij} v_k x^k.\label{MetricTransform}
\end{eqnarray}
Interestingly, we can note that the combination $(-g)^{-1/2}n_{c}/P^{0}$ that appears in the expression 
of the energy-momentum tensor is invariant under such transformations.

It can then be easily checked that the motion equations (\ref{ncevol})-(\ref{Deriv tot P_i 1}) are invariant under the transformations
(\ref{trans dx})-(\ref{CoordinateTransform1})-(\ref{CoordinateTransform2})-(\ref{MetricTransform}). It is also the case for the Einstein equations but the explicit verification is more involved.

The invariance we are  putting forward is clearly a generalization of the extended Galilean invariance. For example, when setting $u_{i}$ to zero, the coordinate transformation reads,
\begin{equation}\label{transunul}
\tx^{i}=x^{i}+d_{i}(\eta),\ \ \ \ \tilde{\eta}=\eta+v_i\,x^{i},
\end{equation}
with $v_{i}=\frac{\dd}{\dd\eta}d_{i}(\eta)$. 
We leave for further studies the exploration of the consequences of such an invariance. In the last section, we will present
a similar symmetry property satisfied  by the motion equations written at subhorizon scales and we will explore in more detail its consequences.

\section{Relativistic streams in the linear regime}

This section aims at highlighting properties of relativistic streams in the linear regime.
From the motion equations (\ref{ncevol}) and (\ref{GeneralPiEvol}), it is straightforward to see that, in a homogeneous metric, the variables $P_i$ and $n_c$ do not depend on time.  Considering that metric inhomogeneities are small compared to background values, we develop a perturbation scheme that consists in expanding each relevant field with respect to 
the metric perturbations\footnote{By definition, $P_{i}^{(0)}=\tau_i$ since $P_{i}^{(0)}$ is a constant and $\tau_i\equiv P_i(\eta_{\rm{in}})$.},
\begin{equation}
P_{i}(\eta,x^{i};\vPi_{i})=\vPi_{i}+P_{i}^{(1)}(\eta,x^{i};\vPi_{i})+P_{i}^{(2)}(\eta,x^{i};\vPi_{i})+\dots
\end{equation}
and
\begin{equation}
n_{c}(\eta,x^{i};\vPi_{i})=n^{(0)}_{c}(\vPi_{i})+n^{(1)}_{c}(\eta,x^{i};\vPi_{i})+n^{(2)}_{c}(\eta,x^{i};\vPi_{i})+\dots \ .
\end{equation}

%
\subsection{A useful property of $P_i^{(1)}$ }

When taking only linear perturbations into account, eq. (\ref{GeneralPiEvol}) reads
\begin{equation}
\dfrac{\dd P_i^{(1)}}{\dd \eta}=\frac{1}{2}\frac{P^{(0)\mu}P^{(0)\nu}}{P^{(0)0}}g^{(1)}_{\mu\nu,i}.
\end{equation}
It appears that $P_i^{(1)}$ is sourced by a gradient term. For scalar adiabatic initial conditions, $P_i^{(1)}$ will therefore remain
a potential field. Note that, as already mentioned in \cite{2014JCAP...01..030D}, this property is a specificity of the variable $P_i^{(1)}$ only. For instance, even at linear order, the momentum field  $P^{i}$ does not derive from a potential. In the particular case of a perturbed Friedmann-Lema\^{i}tre metric, this equation takes the form
\begin{equation}\label{Pi lin 1}
\dfrac{\dd P_i^{(1)}}{\dd \eta}=\partial_i\left(P_j^{(0)}B_j-\dfrac{1}{2}\dfrac{P_j^{(0)}P_k^{(0)}}{P_0^{(0)}}h_{jk}+P_0^{(0)}A\right),
\end{equation}
which allows to express the source term whose $P_i^{(1)}$ is the gradient in terms of the potentials (for example in the conformal Newtonian and synchronous gauges). Note that, to get this expression, only linear terms in the metric perturbations have been taken into account. The same approximation will be used in the rest of this paper since what matters in the nonlinear regime is not metric-metric coupling but the nonlinear coupling between the fields of interest.

\subsection{Linearized equations in Fourier space}

In light of what is done in the study \cite{2014JCAP...01..030D}, we explicitly write in this section the linearized equations in 
Fourier space. To this end, we introduce a Fourier mode $\vk$ and all the fields we study in Fourier space correspond to this mode. We use the same notations as in 
\cite{2014JCAP...01..030D}\footnote{Except $\delta_n$, which applies here to the \textit{comoving} number density whereas it applied to the  \textit{proper} number density in \cite{2014JCAP...01..030D}.},
\begin{eqnarray}\label{Pi}
\vPi_{0}\left(\eta\right)\equiv {P_0}^{(0)} \left(\eta\right)\hbox{,}\
\theta_{\tP}(\vx,\eta)&=&\partial_{i}P_{i}^{(1)}(\vx,\eta), \\ 
P_{i}^{(1)}(\vk,\eta)=\frac{-\ii k_{i}}{k^{2}}\theta_{\tP}(\vk,\eta)\hbox{,}\
\delta_{n}(\vx,\eta)&=&\frac{n_{c}^{(1)}(\vx,\eta) }{n_{c}^{(0)}}.
\end{eqnarray}
Note that  here we have used the fact that $P_{i}^{(1)}(\eta,x^{i})$ is potential 
at linear order. Besides, the on-shell mass constraint imposing
\begin{equation}
P_0^{(1)}=B_i \tau_i +A \tau_0 +\dfrac{\tau_i}{\tau_0} \left(P_i^{(1)}-\dfrac{1}{2}\tau_j h_{ij}\right),
\end{equation}
one has
\begin{equation}
\left(\frac{P^{i}}{P^{0}}\right)^{(1)}=-\frac{P_{i}^{(1)}}{\vPi_{0}}+\frac{P_{j}^{(1)}\vPi_{j}\vPi_{i}}{\vPi_{0}^{3}}
-\frac{\vPi_{i}}{\vPi_{0}}A-B_{i}+\frac{\vPi_{j}h_{ij}}{\vPi_{0}}-\frac{\vPi_{i}}{2\vPi_{0}}\frac{\vPi_{j}\vPi_{k}h_{jk}}{\vPi_{0}^{2}}.
\end{equation}
Thus, after linearizing eq. (\ref{ncevol}), moving to Fourier space and taking the divergence of eq. (\ref{Pi lin 1}), one finally obtains
\begin{equation}\label{Fourier1}
\dot{\delta}_n=i \mu k \dfrac{\tau}{\tau_0}\left(\delta_n+A-h-4\gamma+\kappa(\tau,\tau_0,\mu,h,\gamma)\right)+i k_i B_i +\dfrac{\theta_P}{\tau_0}\left(1-\dfrac{\mu^{2}\tau^{2}}{{\tau_0}^2}\right)
\end{equation}
and
\begin{equation}\label{Fourier2}
\dot{\theta}_P=i \mu k \dfrac{\tau}{\tau_0}\theta_P-k^2\left(\tau_0 A+\tau_j B_j-\tau_0\kappa(\tau,\tau_0,\mu,h,\gamma)\right),
\end{equation}
where $\mu$ is the Cosine of the angle between the wave vector $\vk$ and the initial momentum direction, 
\begin{equation}
\mu=\frac{k_{i}\tau_{i}}{k\tau}\ \ \hbox{with}\ \ \tau^{2}=\tau_{i}^{2},
\end{equation}
where $h$ and $\gamma$ are scalar modes defined so that (see e.g. \cite{Ma:1995ey})
\begin{equation}
h_{ij}=\dfrac{k_i k_j}{k^2}h+\left(\dfrac{k_i k_j}{k^2}-\dfrac{1}{3}\delta_{ij}\right)6 \gamma
\end{equation}
and where
\begin{equation}
\kappa(\tau,\tau_0,\mu,h,\gamma)=\dfrac{\tau^2}{ {\tau_0}^2}\left[\dfrac{1}{2}\mu^2 h+ \gamma \left(3 \mu^2-1\right)\right].
\end{equation}
We can see in particular that the angle between the wave vector $\vk$
and the initial momentum vector plays a significant role in the time evolution of the considered flow.

\medskip

The setting of the adiabatic initial conditions,  essential to numerically solve the system (\ref{Fourier1})-(\ref{Fourier2}), is explained in full detail in appendix \ref{Init} for the particular case of massive neutrinos.

\section{Perturbation Theory with relativistic streams}\label{Relativistic PT}

The aim of this section is to show how the formalism developed to study CDM thanks to Perturbation Theory (PT) can be extended to the study of relativistic species. The regime we will investigate is the one relevant in the context of large-scale structure formation, i.e. we will focus here on subhorizon scales. This restriction will allow us to remove the subdominant coupling terms from the motion equations. It is indeed important to keep in mind that the nonlinear couplings that appear for example in the Einstein equations, in the source term of the Euler equation or in the 
$P^{i}/P^{0}\partial_{i}$ operator do not all have the same amplitude after subhorizon scales  have been reached. 

\subsection{Coupling structure at subhorizon scales}

The identification of the relevant coupling terms is made by comparing the wave number $k$ (characterizing the scale at which the field evolution is studied) with the horizon wave number $k_{\mH}$ (defined as the inverse of the Hubble radius).
First, let us notice that metric perturbations scale like $\delta_{\rho}\, k_{\mH}^{2}/k^{2}$ and that the relative velocity field $(P_{i}-\vPi_{i})$ 
scales like $\delta_{\rho} k_{\mH}/k$, where $\delta_{\rho}$ is the typical energy density contrast. The latter is assumed to be small but can reach values comparable with unity.  This is precisely this regime of PT calculations that we want to explore.

Following the description of CDM fluids at subhorizon scales, in practice we neglect in (\ref{ncevol}) all the terms 
behaving as $k_{\mH}\left({k_{\mH}}/{k}\right)^\alpha$ with $\alpha\geq1$ and in (\ref{Deriv tot P_i 1}) all the terms 
behaving as $k_{\mH}\left({k_{\mH}}/{k}\right)^\alpha$ with $\alpha\geq2$. In this limit, eqs. (\ref{ncevol})-(\ref{Deriv tot P_i 1}) take the form
\begin{eqnarray}
\mD_{\eta}n_{c}+\partial_i(V_{i}n_{c})&=&0\label{Relcons}\\
\mD_{\eta}P_{i}+V_{j}\partial_jP_{i}&=&{\vPi_{0}} \partial_i A + {\vPi_{j}}\partial_i B_j-\frac{1}{2}\frac{\vPi_{j}\vPi_{k}}{\vPi_{0}}\partial_ih_{j k},
\label{RelEuler}
\end{eqnarray}
with
\begin{equation}
\tau_{0}=-\sqrt{m^{2} a^2+\vPi_{i}^{2}},\ \ \mD_{\eta}=\dfrac{\partial}{\partial\eta}-\dfrac{\vPi_{i}}{\vPi_{0}}\dfrac{\partial}{\partial x^{i}},
\end{equation}
and 
\begin{equation}
V_{i}=-\frac{P_{i}-\vPi_{i}}{\vPi_{0}}+\dfrac{\vPi_{i}}{\vPi_{0}}\frac{\vPi_{j}(P_{j}-\vPi_{j})}{(\vPi_{0})^{2}}.\label{Videf}
\end{equation}
The metric perturbations that appear in the Euler equation are computed at linear
order from the Einstein equation. 
Note that in the sub horizon limit, the source term of the Einstein equation is dominated by the fluctuations of the number density.

\subsection{The no-curl theorem and its consequences}

In this paragraph we explicitly demonstrate one of the key results on which rely the carrying out of Perturbation Theory calculations. What we show is that, similarly to the velocity fields of non-relativistic flows, the momentum field $P_{i}$ remains potential to all orders in Perturbation Theory 
(see \cite{2013arXiv1311.2724B,2002PhR...367....1B} and references therein for demonstrations in the non-relativistic case). 

First, let us decompose $P_{i}$ into a potential and a non potential parts,
\begin{equation}
P_{i}=\Phi_{,i}+W_{i},
\end{equation}
with 
\begin{equation}
W_{i,i}=0.
\end{equation}
One can then define the curl field, related to the momentum field via
\begin{equation}
\Omega_{i}=\epsilon_{ijk}P_{j,k}=\epsilon^{ijk}W_{j,k},
\end{equation}
$\epsilon_{ijk}$ being the Levi-Civita symbol, or fully antisymmetric tensor. The objective here is to derive the evolution equation of $\Omega_{i}$ by exploiting the relativistic Euler equation.
Noticing that
\begin{equation}
P_{i,j}=P_{j,i}+\epsilon_{kij}\Omega_{k}
\end{equation}
and applying the operator $\epsilon_{kij}\partial_{k}$ to eq. (\ref{RelEuler}), one obtains 
\begin{equation}
\mD_{\eta}\Omega_{k}+\epsilon_{kij}\epsilon_{mil}\left(V_{l}\,\Omega_{m}\right)_{,j}
+\epsilon_{kij}\left(V_{l}\,P_{l,i}\right)_{,j}=0.
\end{equation}

The last term of the left hand side of this equation eventually vanishes since $V_{l}P_{l,ij}$ is symmetric in $(i,j)$
and
\begin{equation}
V_{l,j}P_{l,i}=-\frac{1}{\vPi^{0}}P_{l,i}P_{l,j}+\frac{1}{(\vPi^{0})^{3}}\vPi_{k}P_{k,j}\,\vPi_{l}P_{l,i},
\end{equation}
which is also symmetric in those indices. So finally we have
\begin{equation}
\mD_{\eta}\Omega_{k}+V_{i}\Omega_{k,i}+V_{i,i}\Omega_{k}-V_{k,i}\Omega_{i}=0,
\end{equation}
which means in particular that the curl field is only sourced by itself. 
Consequently, in the absence of such source terms in the initial conditions, as is the case 
for adiabatic initial conditions, no curl modes will be created in the relativistic flows. 

We are here confronted to a situation very similar to the case of non-relativistic fluids, in which 
curl modes are generated after shell-crossing only. An immediate consequence is that, in standard Perturbation Theory calculations, the evolution equations of the density and $P_{i}$-divergence fields form, together with the equations describing the scalar modes of the metric fluctuations, a complete set of equations.

It is then possible to write those equations on a form easily comparable to the one of a non-relativistic pressureless fluid. To that aim, let us introduce for each fluid labelled by $\vPi_{i}$ the density contrast field $\delta_{\vPi_{i}}$, 
\begin{equation}
\delta_{\vPi_{i}}(\eta,x^{i})=\dfrac{n_c(\eta,x^{i};\vPi_{i})}{n_c^{(0)}(\vPi_{i})}-1.
\end{equation}
The evolution equations then read
\begin{eqnarray}
\mD_{\eta}\delta_{\vPi_{i}}+\left(V_{i}(1+\delta_{\vPi_{i}})\right)_{,i}&=&0\label{Relcons2},\\
\mD_{\eta}P_{i,i}+\left(V_{j}P_{i,j}\right)_{,i}-\mS_{\vPi_{i},ii}&=&0
\label{RelEuler2},
\end{eqnarray}
where $V_{i}$ is related to the field $P_{i}$ via (\ref{Videf}) and where the source term $\mS_{\vPi_{i}}$
is given by
\begin{equation}
\mS_{\vPi_{i}}={\vPi_{0}} A + {\vPi_{j}}B_{j}-\frac{1}{2}\frac{\vPi_{j}\vPi_{k}}{\vPi_{0}} h_{j k}.
\end{equation}
Note that the relation (\ref{Videf}) between $P_{i}$ and $V_{i}$ can be easily inverted,
\begin{equation}
P_{i}=\vPi_{i}\left(1-\frac{1}{1-\vPi_{j}\vPi_{j}/\vPi_{0}^{2}}\ \frac{\vPi_{j}V_{j}}{\vPi_{0}}\right)-\vPi_{0}V_{i}.\label{VidefInv}
\end{equation}
In the following we explore in further detail this system.

\subsection{The extended Galilean invariance}
What are the invariance properties of this system? The original system (\ref{ncevol})-(\ref{Deriv tot P_i 1}) was invariant under
transformations that preserved the operator $P^{\mu}\partial_{\mu}$ present in the left hand side of the equation describing the time evolution of $P_i$. By analogy, here we would like to find transformations that
preserve the operator $\mD_{\eta}+V_{i}\partial_{i}$ present in the left hand side of the corresponding subhorizon equation, while preserving the time variable.
Assuming that $\tau_i$  and $\tau_0$ are unchanged, this can be obtained the following way
\begin{eqnarray}
\tilde x^{i}&=&x^{i}+d_{i}(\eta),\\
\tilde\eta&=&\eta,\\
\tilde\delta_{\vPi_{i}}(\eta,\tilde x^{i})&=&\delta_{\vPi_{i}}(\eta,x^{i}),\\
\tilde V_{i}(\eta,\tilde x^{i})&=&V_{i}(\eta,x^{i})+\partial_{\eta}d_{i}(\eta).
\end{eqnarray}
The transformation rule for the velocity field $V_{i}$ can be re-expressed as a transformation rule for the momentum field $P_{i}$ using (\ref{VidefInv}),
\begin{equation}
\tilde P_{i}(\eta,\tilde x^{i})=P_{i}(\eta,x^{i})
-\vPi_{0}\partial_{\eta}d_{i}(\eta)-\frac{\vPi_{0}}{\vPi_{0}^{2}-\vPi_{j}\vPi_{j}}\ \vPi_{i}\vPi_{j}\partial_{\eta}d_{j}(\eta).
\end{equation}
$P_{i,i}$ and all the potentials being unchanged under such transformations, the invariance of the equation of interest is ensured.
This result is an extension of the extended Galilean invariance satisfied by the CDM flow. For CDM, as it has been stressed in recent papers (\cite{2013NuPhB.873..514K,2013JCAP...05..031P}), this property has important consequences regarding large-scale structure formation. In particular, one expects the unequal time correlation functions of fields of that type to obey Ward identities. We will derive them for relativistic species in terms of the power spectra of the Fourier modes.

\subsection{Nonlinear equations in Fourier space}

We complete this work by presenting a global motion equation in Fourier space, showing explicitly the coupling structure of the motion equations. So let us introduce
the velocity divergence field in units of - $\mH$,  $\theta_{\vPi_{i}}(\eta,x^{i})$\footnote{In the non-relativistic limit, $\partial_i P_i=-m a \partial_i V^i$.},
\begin{equation}
\theta_{\vPi_{i}}(\eta,x^{i})=-\frac{P_{i,i}(\eta,x^{i};\vPi_{i})}{m a\mH}.
\end{equation}
When written in Fourier space, eqs. (\ref{Relcons2})-(\ref{RelEuler2}) read
\begin{eqnarray}
\left(a\partial_a-\ii\dfrac{\mu k \tau }{\mH \tau_0}\right)
\delta_{\vPi_{i}}(\vk)
+\frac{m a}{\vPi_{0}}\left(1-\dfrac{\mu^{2}\tau^2}{\tau_0^2}\right)\theta_{\vPi_{i}}(\vk)
&=&\nonumber\\
&&\hspace{-3cm}-\dfrac{m a}{\tau_0}\int{\dd^3\vk_{1} \dd^3\vk_{2}}\alpha_{R}(\vk_{1},\vk_{2};\vPi_{i})\delta_{\vPi_{i}}(\vk_{1})\theta_{\vPi_{i}}(\vk_{2})
\label{Fouriersub1}\\
\left(1+a \dfrac{\partial_a \mH}{\mH}+a \partial_a-\ii \dfrac{\mu k \tau}{\mH\tau_0 }\right)\theta_{\vPi_{i}}(\vk)-\dfrac{k^2}{m a\mH^2}\mS_{\vPi_{i}}(\vk)&=&
\nonumber\\
&&\hspace{-3cm}-\dfrac{m a}{\tau_0}\int{\dd^3\vk_{1} \dd^3\vk_{2}}
\beta_{R}(\vk_{1},\vk_{2};\vPi_{i})\theta_ {\vPi_{i}}(\vk_{1})\theta_ {\vPi_{i}}(\vk_{2}),
\label{Fouriersub2}
\end{eqnarray}
where $\mS_{\vPi_{i}}(\vk)$ is the Fourier transform of the field $\mS_{\vPi_{i}}(\vx)$,
\begin{equation}
\mS_{\vPi_{i}}(\vk)=\tau_0 A(\vk)+\vec{\tau}\cdot\vec{B}(\vk)-\frac{1}{2}\frac{\vPi_{i}\vPi_{j}}{\vPi_{0}}h_{ij}(\vk)
\end{equation}
and where the kernel functions are defined as
\begin{align}\label{alphaR}
\alpha_{R}(\vk_{1},\vk_{2};\vPi)=\Dirac(\vk-\vk_{1}-\vk_{2})\dfrac{(\vk_{1}+\vk_{2})}{k_2^2}\cdot\left[\vk_{2}-\vec{\tau}\dfrac{\vk_{2}\cdot \vec{\tau}}{\tau_0^2}\right],
\end{align}
\begin{align}\label{betaR}
\beta_{R}(\vk_{1},\vk_{2};\vPi)=\Dirac(\vk-\vk_{1}-\vk_{2})\dfrac{\left(\vk_{1}+\vk_{2}\right)^{2}}{2k_1^2 k_2^2}\left[\vk_{1} \cdot \vk_{2}-\dfrac{\vk_{1}\cdot \vec{\tau}\vk_{2}\cdot \vec{\tau}}{\tau_0^2}\right].
\end{align}
A remarkable property is that the kernel functions
$\alpha_{R}$  and $\beta_{R}$ depend on the flow considered via the variable  $\vec{\tau}$. In the non-relativistic limit (i.e. when $\tau_0 \rightarrow -ma $ and $\tau_i \rightarrow 0$), we recover the standard equations and the kernel functions  that appear in the CDM flow equation (see \cite{2012PhRvD..85f3509B}).

We are now in position to write the full equation of motion, including the scale-scale nonlinear couplings 
in the presence of cold and hot non-interacting dark matter. We recall that these equations are valid until the first shell-crossing occurs.
Formally, we consider a collection of $n$ streams. Each stream is single-flow. It corresponds to either a CDM component or  a baryonic component or
a massive neutrino component. All these fluids obey the very same motion equations so there is no point in the following to distinguish 
them from one another.

It means that the time-dependent $2n$-uplet,
\begin{equation}
\Psi_{a}(\vk)=(\delta_{\vPi_1}(\vk),\theta_{\vPi_{1}}(\vk),\dots,\delta_{\vPi_{n}}(\vk),\theta_{\vPi_{n}}(\vk))^{T},
\end{equation}
contains all the relevant field components. Note that the Einstein equations relate the potentials to those fields
thus these potentials are eventually  eliminated.

In this context, the motion equations (\ref{Fouriersub1}) and (\ref{Fouriersub2}) can formally be recast in the form
\begin{equation}
\partial_{\eta}\Psi_{a}(\vk)+\Omega_{a}^{\ b}\,\Psi_{b}(\vk)=\gamma_{a}^{\ bc}(\vk_{1},\vk_{2})\Psi_{b}(\vk_{1})\Psi_{c}(\vk_{2}),
\label{FullEoM}
\end{equation}
where the indices $a$ and  $b$ run from 1 to $2n$\footnote{The Einstein notation for the summation over repeated indices is adopted.}. In the right hand side of this equation, it is
assumed that the wave modes are integrated over.
The matrix elements $\Omega_a^{\ b}$ encode the linear theory couplings. They contain in particular the way in which the source terms $\mS_{\vPi_{i}}(\vk)$ 
can be re-expressed as a function of the $2n$-uplet elements.

Besides, the \emph{symmetrized vertex} matrix $\gamma_{a}^{\ bc}(\vk_{1},\vk_{2})$ describes the nonlinear 
interactions between different Fourier modes. Its components are given by
\begin{eqnarray}
\gamma_{2p-1}^{\ 2p-1\,2p}(\vk_1,\vk_2)&=&-\frac{ma}{2\vPi_{0}}
\alpha_{R}(\vk_{1},\vk_{2},\vPi_{p})\\
\gamma_{2p}^{\ 2p\,2p}(\vk_1,\vk_2)&=&
-\frac{ma}{\vPi_{0}}
\beta_{R}(\vk_{1},\vk_{2},\vPi_{p}),
\label{vertexdefinition}
\end{eqnarray}
with $\gamma_{a}^{\ bc}(\vk_1,\vk_2)=\gamma_{a}^{\ cb}(\vk_2,\vk_1)$ and $\gamma_{a}^{\ bc}=0$ 
otherwise. Contrarily to the pure CDM case, the $\gamma_{a}^{\ bc}$ matrix elements depend
on time (and on the background evolution) for each mode through the time evolution of $\vPi_{0}$. 
Remarkably though, they encode all the nonlinear couplings of the system, which is formally similar to that
of a multi-component system of pressureless fluids.

Equation (\ref{FullEoM}) is the main result of this paper. It encodes the evolution of streams of relativistic or non-relativistic particles 
in the nonlinear regime at subhorizon scales. It can in particular be used in the context of the growth of large-scale structure in presence of massive neutrinos.
At this stage however we do not propose an operational procedure to implement such Perturbation Theory calculations. To do so, one could think for example  about making use of the so-called Time Renormalization Group (TRG) approach introduced in \cite{2008JCAP...10..036P}. 
Indeed, thanks to the motion equation (\ref{FullEoM}), it is possible to compute the time derivative of products such as  $\Psi_{a}(\vk,\eta)\Psi_{b}(\vk',\eta)$
or  $\Psi_{a}(\vk_{1},\eta)\Psi_{b}(\vk_{2},\eta)\Psi_{c}(\vk_{3},\eta)$.
Once their ensemble averages computed, one can get the coupled evolution equations for the power spectra (see \cite{2008JCAP...10..036P}).
Provided the truncation is properly made, such equations encompass the standard Perturbation Theory calculations but with the advantage that no explicit computation of the linear Green function is necessary.  The simplicity of this approach has already been advocated in this context 
in \cite{2009JCAP...06..017L}, where it is used to evaluate the impact of massive neutrinos on structure growth (but restricting the neutrino fluid 
to its linear behavior).

\subsection{The Ward identities}

Let us define unequal time correlators as
$\langle\delta_{\vPi_{1}}(\eta_{1},\vk_{1})\dots
\delta_{\vPi_{n}}(\eta_{n},\vk_{n})\rangle$
for a collection of flows $\vPi_{i}$. Due to statistical homogeneity, such quantities 
are expected to be proportional to $\Dirac\left(\sum_{i}\vk_{i}\right)$. One can then define the multi-point power spectra
$\mP$ so that
\begin{equation}
\langle\delta_{\vPi_{1}}(\eta_{1},\vk_{1})\dots
\delta_{\vPi_{n}}(\eta_{n},\vk_{n})\rangle=\Dirac\left(\sum_{i}\vk_{i}\right)\mP_{\vPi_{1},\dots,\vPi_{n}}
\left(\eta_{1},\vk_{1},\dots,\eta_{n},\vk_{n}\right).
\end{equation}
Following \cite{2013NuPhB.873..514K}, \cite{2013JCAP...05..031P} and \cite{2014JCAP...04..011P}, one can derive
Ward identities that give consistency relations between those quantities.

We denote $\tilde\delta(\vk,\eta)$ the Fourier density contrast in presence of a large-scale displacement field $d_{i}(\eta)$ (with an arbitrary time dependence and $d_i$ being treated linearly).
It can be expressed as a function of the Fourier density contrast in absence of such displacement thanks to a simple phase shift,
\begin{equation}
\tilde\delta_{\vPi_{i}}(\vk,\eta)=\exp(\ii\vk.\vd)\delta_{\vPi_{i}}(\vk,\eta)\approx (1+\ii\vk.\vd)\delta_{\vPi_{i}}(\vk,\eta).
\label{Wardddep}
\end{equation}
 This relation gives explicitly the dependence of each mode  on a large-scale displacement field. The Ward identities 
 are then obtained by relating such large displacement fields to long-wave modes. More precisely, one can define adiabatic modes inducing equal displacements in all the flows, denoted $\delta_{\rm adiab.}(\vq)$, and 
satisfying
\begin{equation}
\vd_{\rm adiab}(\eta,\vx)=\int{\dd^{3}\vq\frac{-\ii\vq}{q^{2}}e^{\ii\vq.\vx}\delta_{\rm adiab.}(\eta,\vq)}.
\end{equation}
This definition imposes\footnote{To get this result, we have neglected the $e^{-\ii\vq.\vx}$ term because we are interested here in large wavelenghts.}
\begin{equation}
\langle
\delta_{\rm adiab.}(\eta,\vq)\vd_{\rm adiab}(\eta',\vx)
\rangle
=\frac{\ii \vq}{q^{2}}\,\mP_{\rm adiab.}(\eta,\eta',q),
\end{equation}
$\mP_{\rm adiab.}(\eta,\eta',q)$ being the unequal time power spectrum of adiabatic modes.
Making use of eq. (\ref{Wardddep}), the correlator reads for an adiabatic displacement
\begin{equation}\label{corr}
\langle\tilde\delta_{\vPi_{1}}(\eta_{1},\vk_{1})\dots
\tilde\delta_{\vPi_{n}}(\eta_{n},\vk_{n})\rangle=
(1+\ii\sum_{i}\vk_{i}.\vd_{\rm adiab.}(\eta_{i}))\langle\delta_{\vPi_{1}}(\eta_{1},\vk_{1})\dots
\delta_{\vPi_{n}}(\eta_{n},\vk_{n})\rangle,
\end{equation}
where the ensemble average is performed over all the modes except those participating in the large-scale displacement perturbation.
Finally, assuming that the only dependence with a large-scale adiabatic mode is in the displacement field, 
one can eventually derive the following relation,
\begin{eqnarray}
\mP_{{\rm adiab.},\vPi_{1},\dots,\vPi_{n}}(\eta,\vq,\eta_{1},\vk_{1},\dots,\eta_{n},\vk_{n})
&=&\\\nonumber
&&\hspace{-4cm}-\sum_{i}\frac{\vk_{i}.\vq}{q^{2}}\mP_{\rm adiab.}(\eta,\eta_{i},q)\  
\mP_{\vPi_{1},\dots,\vPi_{n}}(\eta_{1},\vk_{1},\dots,\eta_{n},\vk_{n}).
\label{wardidentities}
\end{eqnarray}
It is obtained by computing the average of the product between the quantity at play in eq. (\ref{corr}) and $\delta_{\rm adiab.}(\eta,\vq)$. This relation is valid for $q\ll k_{i}$.
Note that the right hand side of the relation (\ref{wardidentities}) automatically vanishes when all the time variables 
are equal as, in that case, one expects the result to be proportional to  $\Dirac\left(\sum_{i}\vk_{i}\right)$.

\section{Conclusions and perspectives}

We have presented a derivation of fully nonlinear evolution equations for  streams of relativistic particles in an arbitrary background. 
The derivation of these equations is entirely based on conservation laws. They lead to the equations (\ref{ncevol}) and (\ref{GeneralPiEvol}), 
which form a closed system once the background is given.

The key point allowing to make this construction sensible is the fact that fluids of non-interacting particles, such as neutrinos, can be decomposed into a collection of streams, each of them obeying the independent motion equations we derived. 
This is the essence of eq. (\ref{mapping1}). It has been explicitly shown in
\cite{2014JCAP...01..030D} that this decomposition is effective at the level of the linear evolution of the whole neutrino fluid.
The initial number density of particles in each stream can be computed once initial conditions and gauge are specified. In \cite{2014JCAP...01..030D}, we computed them for adiabatic initial conditions in the conformal Newtonian gauge. Here we extend 
the results to the synchronous gauge in order to be more exhaustive.

The last section of this study is devoted to the exploration of the coupling structure that appears once the motion equations are restricted to subhorizon scales. 
In this derivation, we retained only dominant nonlinear coupling terms based on a power counting argument.
The resulting equations, (\ref{Relcons})-(\ref{RelEuler}), appear as a slight extension of those describing flows of cold dark matter at subhorizon scales. However, we think they capture all the relevant nonlinear couplings. The exploration of the properties of the resulting system gives very promising insights. We recall here the two most important points we noticed. 
The first one is that the momentum field $P_{i}$ remains potential even in the nonlinear regime. It implies that, similarly to non-relativistic ones, relativistic streams can entirely be described  by introducing a two-component scalar doublet containing the number density of particles and their velocity divergence. The second key element is that the couplings are only quadratic in the fields\footnote{This is true at subhorizon scales only.}, as for CDM. As a result the overall motion equation, which takes into account all the 
streams, can be recast in the formal form (\ref{FullEoM}). This is the main result of this paper. It provides a starting point for the implementation of Perturbation Theory calculations involving relativistic species, such as neutrinos.

Note also that, throughout the paper, we paid attention to the invariances properties of the systems we studied. In particular, we showed  that eq. (\ref{FullEoM})
satisfies an extended Galilean invariance. Interestingly, it paves the way for a further exploration of the mode coupling structure, and particularly for a description of how the long-wave modes and the short-wave modes interact. We expect in particular that the relative motions that exist between the different streams act as a particularly efficient coupling mechanism. Indeed, as shown in \cite{2010PhRvD..82h3520T}, it is already the case for baryon-CDM mixtures. An effective way to address this issue is to exploit the  eikonal approximation, as presented in Ref. \cite{2012PhRvD..85f3509B,2013PhRvD..87d3530B}. 
We leave for further studies those calculations.

\textbf{Acknowledgements}:  This work is partially supported by the grant ANR-12-BS05-0002 of the French Agence Nationale de la Recherche.
\appendix
\section{Adiabatic initial conditions for massive neutrinos}\label{Init}

We revisit here the setting of the initial conditions as presented in \cite{2014JCAP...01..030D}: at initial time we assign to the flow
labelled by $\tau_i$ all the neutrinos whose momentum $P_{i}$ is equal to $\vPi_{i}$ within $\dd^{3}\vPi_{i}$. The initial time $\eta_{\rm{in}}$ is chosen so that the neutrino decoupling occurs at a time $\eta<\eta_{\rm{in}}$ 
and neutrinos become non-relativistic at a time $\eta>\eta_{\rm{in}}$. The solutions we describe correspond to adiabatic initial conditions. We choose the simplest alternative respecting the adiabaticity constraint, i.e.
\begin{equation}
P_{i}(\etain,\vx;\vPi_{i})=\vPi_i.
\end{equation}
It implies in particular that  ${P_{i}}^{(1)}(\vx,\etain;\vPi_{i})=0$ and consequently that 
\begin{equation}
\theta_P(\vx,\etain;\vPi_{i})=0.
\end{equation}

Besides, before decoupling, the background distribution of neutrinos is expected to follow a Fermi-Dirac law $f_0$ with a given temperature $T$ and no chemical potential (see e.g. Refs. \cite{2013neco.book.....L,Lesgourgues2006,1994ApJ...429...22M,Ma:1995ey} for a physical justification of this assumption). Then, 
as explained in Ref. \cite{2013neco.book.....L}, after neutrino decoupling the phase-space distribution function of neutrinos 
is still a Fermi-Dirac distribution, that we express here in terms of the momentum $q$ defined so that the energy measured by an observer at rest 
in the metric, $\epsilon$, satisfies 
\begin{equation}
\epsilon^{2}=m^{2}+(q/a)^{2}.
\end{equation}
Nonetheless, after decoupling, the temperature is expected to vary locally, whence
\begin{equation}\label{fform}
f\left(\etain,\vx,q\right)\propto \frac{1}{1+\exp\left[q/(a k_B (T+\delta T(\etain,\vx))\right]},
\end{equation}
$k_B$ being the Boltzmann constant. The relation between the energy (and thus $q^\mu$) and $p^\mu$ as well as the expression of $\delta T / T$ in terms of the metric perturbations depend on the gauge chosen. 

In the synchronous and conformal Newtonian gauges, the momentum variable is defined so that $\epsilon=-U^\mu p_\mu=-U^0 p_0$, $U^\mu$ being the four-velocity of the comoving observer. Thus 
\begin{align}
p_0=-a \epsilon (1+A).
\end{align}
Eq. (\ref{fform}) can thereby be re-expressed in terms of the variable $p_{i}$ thanks to  the relation 
\begin{equation}
q=\tau-\dfrac{\tau_0^2}{\tau}\kappa(\tau,\tau_0,\mu,h,\gamma)+\dfrac{\tau_j }{\tau} p_j^{(1)},
\end{equation}
which gives
\begin{equation}
f\left(\eta_{\text{in}},\mathbf{x},p_j\right)\propto \left(1+\exp\left[\dfrac{\tau-\dfrac{\tau_0^2}{\tau}\kappa(\tau,\tau_0,\mu,h,\gamma)+\dfrac{\tau_j }{\tau} p_j^{(1)}}{a k_B (T+\delta T(\eta_{\text{in}},\mathbf{x}))}\right]\right)^{-1}.
\end{equation}
One can see in particular that 
\begin{equation}
f^{(1)}\left(\eta_{\text{in}},\mathbf{x},p_j\right)=-\left(\dfrac{\delta T}{T}+\dfrac{\tau_0^2}{\tau^2}\kappa(\tau,\tau_0,\mu,h,\gamma)\right)\dfrac{\dd f_0(p)}{\dd \log p}.
\end{equation}
In the conformal Newtonian gauge, we recover of course eq. (4.19) of \cite{2014JCAP...01..030D}.
The second initial condition we need is therefore
\begin{equation}
\delta_{n}(\etain,\vx;\vPi_{i})=-\left(\dfrac{\delta T}{T}+\dfrac{\tau_0^2}{\tau^2}\kappa(\tau,\tau_0,\mu,h,\gamma)\right)\frac{\dd\log f_{0}(\vPi)}{\dd\log \vPi}.
\end{equation}

As mentioned in \cite{Lesgourgues2006}, on super-Hubble scales, the temperature perturbation of the neutrino fluid is proportional to its density contrast: ${4\delta T(\vx,\etain)}/{T(\etain)}= \rho^{(1)}(\vx,\etain)/\rho^{(0)}(\etain)$. Besides, the adiabaticity hypothesis imposes equality between the initial density contrasts of all species. Using the standard results that, for photons, $\rho_\gamma^{(1)}(\vx,\etain)/\rho_\gamma^{(0)}(\etain)=-2{\psi(\vx,\etain)}$ in the conformal Newtonian gauge and $-\dfrac{2}{3} h_{ii}$ in the Synchronous gauge, one thus finds
\begin{equation}
{\delta T(\vx,\etain)}/{T(\etain)}=-{\psi(\vx,\etain)}/{2} \\\ \hbox{ in the conformal Newtonian gauge }
\end{equation}
and
\begin{equation}
{\delta T(\vx,\etain)}/{T(\etain)}=-{h_{ii}(\vx,\etain)}/{6} \\\ \hbox{ in the Synchronous gauge }.
\end{equation}
These relations are useful in order to implement the numerical resolution of the linearized motion equations, as presented in detail in 
 \cite{2014JCAP...01..030D}.

\bibliographystyle{JHEP}
\bibliography{neutrinos,LesHouches}
\end{document}